\documentclass[prb,twocolumn,showpacs,preprintnumbers,amsmath,amssymb]{revtex4}
\usepackage{graphicx}% Include figure files
\usepackage{dcolumn}% Align table columns on decimal point
\usepackage{bm}% bold math

\begin{document}

%\preprint{S.X. Dou et al./Superconductivity...}

\title{Superconductivity, critical current density and flux pinning in MgB$_{2-x}$(SiC)$_{x/2}$ superconductor after SiC nanoparticle doping}

\author{S. X. Dou}
\author{A. V. Pan}
%\email{pan@uow.edu.au}
\author{S. Zhou}
\author{M. Ionescu}
\author{X. L. Wang}
\author{J. Horvat}
\author{H. K. Liu}
\affiliation{Institute for Superconducting and Electronic Materials, University of Wollongong, Wollongong, NSW 2522, Australia}
\author{P. R. Munroe}
\affiliation{Electron Microscopy Unit, University of New South Wales, Sydney, NSW 2052, Australia}

\date{July 1, 2002} 

\begin{abstract}
We investigated the effect of SiC nano-particle doping on the crystal lattice structure, critical temperature $T_c$, critical current density $J_c$, and flux pinning in MgB$_2$ superconductor. A series of MgB$_{2-x}$(SiC)$_{x/2}$ samples with $x = 0$ to 1.0 were fabricated using in-situ reaction process. The contraction of the lattice and depression of $T_c$ with increasing SiC doping level remained rather small due to the counter-balanced effect of Si and C co-doping. The high level Si and C co-doping allowed the creation of intra-grain defects and highly dispersed nano-inclusions within the grains which can act as effective pinning centers for vortices, improving $J_c$ behavior as a function of the applied magnetic field. The enhanced pinning is mainly attributable to the substitution-induced defects and a local structure fluctuations within grains. A pinning mechanism is proposed to account for different contributions of different defects in MgB$_{2-x}$(SiC)$_{x/2}$ superconductors.
\end{abstract}

\pacs{74.62.Dh, 74.62.Bf, 74.60.Ge}

\maketitle

\section{Introduction}

Extensive research efforts have been made for increasing the critical temperature ($T_c$) in MgB$_2$ since superconductivity in this compound was discovered \cite{nature1}. The most commonly used procedures are high pressure treatment and element substitutions as summarized in a recent review article by Buzea and Yamashita \cite{review} and references therein. It appears that all element substitutions caused crystal lattice contraction and reduced $T_c$. The elements which do not considerably reduce $T_c$ are Zn, Si and Li. However, all element substitutions reported so far are limited to single element doping. Because of the rigid lattice structure and small number of elements in the MgB$_2$ compound the level of substitution is substantially restricted, otherwise $T_c$ would be strongly depressed. Furthermore, the majority of studies on substitution was carried out to replace Mg except recent work on C \cite{takenobu,mickelson} and Si \cite{cimberle} substitutions aimed at replacing B.

Another important goal of substitution is enhancement of pinning in MgB$_2$. In spite of impressive progress on improving the critical current density ($J_c$) \cite{goldacher,suo,grasso,glowacki,jin,soltan,takano} at temperatures above 20~K, which is considered to be the benchmark operating temperature for this material, $J_c$ rapidly drops with increasing magnetic field ($H$) due to its poor pinning ability. If MgB$_2$ is to be useful in high fields, the flux pinning strength must be improved. Attempts have been made to enhance the flux pinning by oxygen alloying in MgB$_2$ thin films \cite{eom} and by proton irradiation of MgB$_2$ powder \cite{bogu}, which led to an encouraging improvement of both irreversibility field ($H_{\rm irr}$) and $J_c(H)$. The question is whether one can introduce effective pinning centers into MgB$_2$ by chemical doping. Numerous attempts have been made to improve flux pinning using this procedure. The results for doping into MgB$_2$ reported so far are largely limited to addition rather than substitution of the doping elements. Additives appeared to be ineffective for pinning enhancement at high temperatures \cite{zhao,feng,wang}. However, if doping elements are substituted into, instead added to MgB$_2$ crystals, pinning properties can be much improved by induction of a net of intra-granular defects due to crystal lattice distortions and local fluctuations of the superconducting order parameter. Both can strongly contribute to pinning. Indeed, pinning reinforcement research in MgB$_2$ has to be focused on intra-granular pinning as the weak-link problem is insignificant for this material \cite{nature2,dou}, unlike for high-$T_c$ systems. Therefore, it is necessary to have a high level and multi-element substitution, creating effective pinning sites but not lowering $T_c$ dramatically. Indeed, for MgB$_2$, the boron plane is responsible for superconductivity, therefore it is desirable for substitution to take place of B positions, which would induce local fluctuations of the superconducting order parameter.

In our previous work \cite{sic}, we doped MgB$_2$ with SiC nano-particle powder, producing MgB$_2$(SiC)$_x$ compound where $x = 0$ to 0.34. It was found that the added SiC was {\it dissolved} within the crystal lattice; that Si and C co-substitution resulted in a surprisingly modest reduction of $T_c$ (only 2.6~K for $x = 0.34$); and that vortex pinning was significantly increased, drastically improving $J_c(H)$ behavior \cite{sic}. However, from the doping procedure it was difficult to estimate the level of substitution upon increasing $x$. Excessive secondary phases also appeared, as 10~nm inclusions inside the MgB$_2$ grains. It was not quite clear what was the predominating factor for the enhanced pinning: the substitution or the inclusions. In this detailed work, we report not only the influence of SiC co-substitution for B in the form of MgB$_{2-x}$(SiC)$_{x/2}$ with $x$ varying from 0 to 1 on different aspects of superconductivity in its material, but also show that the substitution governs the improved $J_c(H)$ performance, whereas impurities (inclusions) bear an assisting role, having a considerable contribution to the pinning of vortices presumably at relatively low fields.

In  the next section we describe the sample preparation procedure and experimental details. In Section \ref{disc}, we show the effect of substitution on MgB$_2$ crystal lattice contraction and phase composition (Sect.~\ref{crlat}); in Section \ref{vol} we discuss reasons for a small critical temperature degradation for relatively high SiC-doping level, and show that $T_c$ degradation due to the substitution is similar to the effect of hydrostatic pressure applied to MgB$_2$; in Section \ref{j} we mainly deal with critical current density behavior as a function of the applied field and temperature, providing evidence for the improved performance as a result of SiC doping procedure. We also propose a quantitative pinning mechanism to account for the $J_c(H)$ behavior. In the last section we summarize the obtained results and conclusion. 

\section{EXPERIMETANL}

MgB$_2$ samples were prepared by the in-situ reaction, which has been previously described in details \cite{wang2}. Powders of magnesium (99\%) and amorphous boron (99\%) were well mixed with powder of SiC nano-particles, varying from 10~nm to 100~nm in size, with an atomic ratio of MgB$_{2-x}$(SiC)$_{x/2}$,  where $x = 0$, 0.04, 0.1, 0.2, 0.3, 0.5, 0.65, 0.8, and 1.0. The mixed powders were filled into Fe tubes. The composite tubes were groove-rolled, sealed in a Fe tube and then directly heated at preset temperatures to $950^{\circ}$C, for 1 hour in flowing high purity Ar. This was followed by quenching to the liquid nitrogen temperature.

X-ray-powder diffraction (XRD) measurements were performed on undoped and SiC-doped samples. The X-ray scans were recorded using Cu$_{{\rm K} \alpha} = 1.5418$~{\AA}, and indexed within the space group P6/mmm. Microstructure and micro-composition were characterized using scanning electron microscope (SEM) and transmission electron microscope (TEM) equipped with an energy dispersive X-ray spectroscopy analyzer (EDS) and mapping analysis. The magnetization as a function of temperature $T$ and magnetic field $H$ applied along the longest sample dimension was measured using a Quantum Design Magnetic Property and Physical Property Measurement Systems within the field range $|H| \le 9$~T, and within the temperature range of $5~{\rm K} \le T \le 30$~K.  A magnetic $J_c$ was derived from the half-width of magnetization loops $\Delta M = (|M^+| + |M^-|)/2$ ($M^+$ and $M^-$ are descending and ascending branches of the magnetization loop, respectively), using the following critical state model formula: $J_c = k10 \Delta M/d$, where $k = 12w/(3w-d)$ is a geometrical factor, with $d$ and $w$ being sample thickness and width, respectively. All the samples were of a similar rectangular shape and size. Because of the strong thermo-magnetic flux-jump instabilities the low field $J_c$ could not be measured at $T \le 10$~K.

\section{RESULTS AND DISCUSSION}
\label{disc}

\subsection{Structural change of MgB$_2$ due to substitution}
\label{crlat}

\begin{figure}
\vspace{-0.5cm}
\includegraphics[scale=0.47]{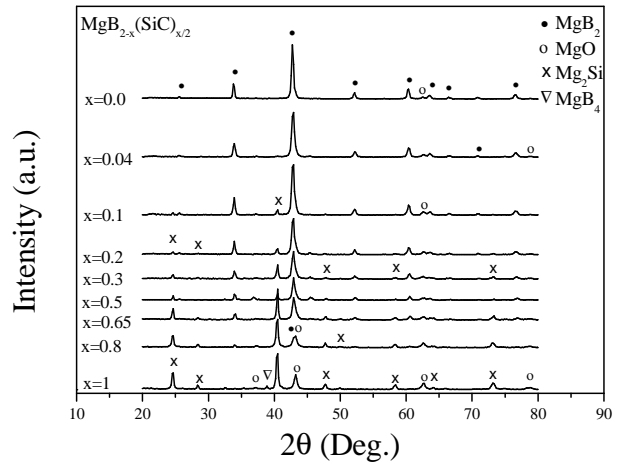}
\vspace{-6.0cm}
\caption{\label{xrd}X-ray diffraction patterns for all the measured MgB$_2$(SiC)$_{x/2}$ samples.}
\end{figure}

\begin{figure}
%\vspace{-0.5cm}
\includegraphics[scale=0.3]{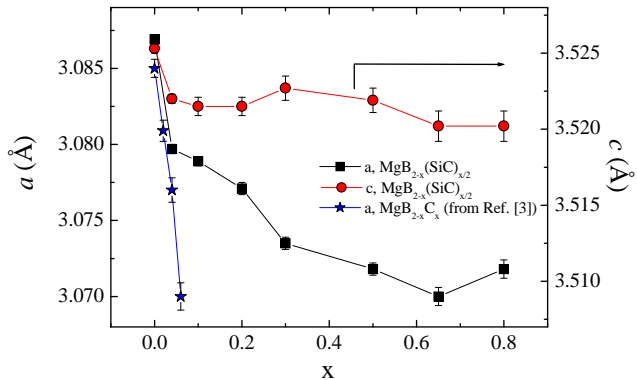}
\caption{\label{cl}Lattice parameters $a$ and $c$ as a function of the nominal SiC content $x$. For comparison, the variation of the $a$-axis for the single element C substitution for B taken from Ref.~\onlinecite{takenobu} is also shown.}
\end{figure}

Fig.~\ref{xrd} shows X-ray diffraction patterns for the series of MgB$_{2-x}$(SiC)$_{x/2}$ samples where $x = 0$ to 1.0. For the in-phase reflection which occurs between $2\theta = 33^{\circ}$ and $2\theta = 34^{\circ}$ indexed as (100), the centroid of the peak clearly shifts to higher $2\theta$ values with increasing $x$, indicating a contraction of the $a$-axis of the crystal lattice. For the (002) reflection peak, which occurs between $2\theta = 51^{\circ}$ and $2\theta = 52^{\circ}$, the shift to higher $2\theta$ values with increasing $x$ is relatively small. The change of the lattice parameters, $a$ and $c$ of the hexagonal AlB$_2$-type structure of MgB$_2$ were calculated using the peak shifts shown in Fig.~\ref{cl}.

At lower doping level ($x < 0.1$), the sample consists of a major phase with MgB$_2$ structure. At $x \ge 0.1$ minority phase Mg$_2$Si and trace amount of MgO were identified. The amount of these non-superconducting phases was increased with increasing SiC doping level. At $x = 0.1$ to 0.2, the minority phase Mg$_2$Si occupied about 10\% volume fraction. This minority phase increased to about 30\% volume fraction for $0.3 \le x \le 0.5$. At $x =0.65$, the amount of non-superconducting phases {\em already exceeded} MgB$_2$. At $x = 1.0$, the reflection peaks (100) and (002) for MgB$_2$ structure disappeared, and the sample consists of multi-phases which can no longer be indexed as the MgB$_2$ structure (Fig.~\ref{xrd}).

Both crystal lattice axes are sharply reduced at low doping levels $x \le 0.04$ (Fig.~\ref{cl}), nevertheless the sample overwhelmingly remained as the single phase (Fig.~\ref{xrd}). After the sharp contraction, the $a$-axis continues to moderately decrease with increasing doping level till the nominal composition $x = 0.65$ is reached. In the same time, the $c$-axis remains almost constant. The negligible $c$-axis variations are within the experimental error. These results suggest that at the low doping level the SiC-dopant is to a large extent dissolved in the MgB$_2$ crystal lattice. As the doping level grows the substitution effect slows down, but it can still be quite effective up to $x \simeq 0.5$, where the $a$-axis change reaches a quasi-plateau (Fig.~\ref{cl}). Furthermore, the slow variation in the $a$-axis with increasing SiC doping level indicates that B was substituted by C and Si. In comparison, pure C substitution for B resulted in a reduction of the $a$-axis from 3.085 to 3.070 at $x = 0.065$ \cite{takenobu} (Fig.~\ref{cl}), while the co-doping of Si and C for B reached the similar level of the $a$-axis reduction at the nominal composition $x = 0.65$. This is an order magnitude higher than the single C doping (Fig.~\ref{cl}). It is clear that the Si and C substitution raised the saturation level considerably. It is probably due to the fact that the average atomic radius of C (0.91~\AA) and Si (1.46~\AA) closely matches that of B (1.17~\AA). Therefore, this co-doping counter-balanced the negative effect of the single element doping \cite{sic}. The formation of Mg$_2$Si phase could not consume all of Si, the crystal lattice would then show a similar contraction to the one obtained for the single element C-doping \cite{takenobu} (see Fig.~\ref{cl}) accompanied by the strong decrease of $T_c$ \cite{takenobu}. This was not observed in our experiment (see Sect.~\ref{vol}).

\begin{figure}
%\vspace{-0.5cm}
\includegraphics[scale=0.4]{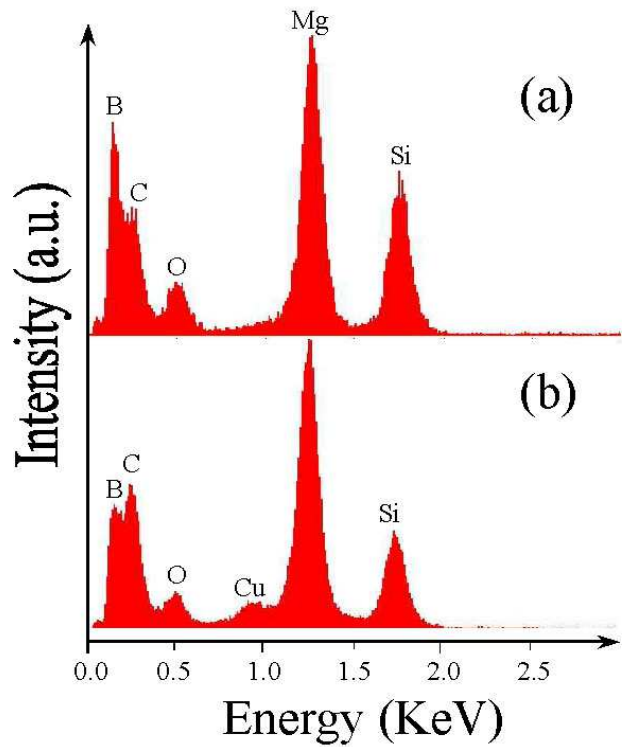}
\caption{\label{edx}The EDS analysis results taken on single grains of $x = 0.2$ (a) and $x = 0.5$ (b) samples. It shows the co-existence of Si, C, B, and Mg within each single grain. A small Cu-peak in (b) appeared due to the background signal from the sample holder.}
\end{figure}

The simultaneous incorporation of C and Si into the lattice structure is also supported by energy dispersive X-ray spectroscopy analysis on single grains of MgB$_2$. The analysis clearly shows that Si and C co-exist with Mg and B within the same grain (Fig.~\ref{edx}). Moreover, the $({\rm Si}+{\rm C})/{\rm B}$ ratio obtained for samples having different doping levels quantitatively correlates with the corresponding doping level of the sample. However, we observed local variations of the signal intensity for C and Si across each analyzed grain, suggesting very local variations in the substitution level and making a qualitative assessment impossible. This is in contrast to Mg and B, which appear to be rather uniformly distributed across each analyzed grain. In Fig.~\ref{edx}, we present the spectral analysis for two samples having $x = 0.2$ (a) and $x = 0.5$ (b). Note that the B-peak intensity became comparable to the intensities of Si- and C-peaks for the higher doping level sample (Fig.~\ref{edx}(b)), pointing out that Si and C replace B within the MgB$_2$ grains.

The spectral analysis clearly shows an O peak within the MgB$_2$ grains (Fig.~\ref{edx}). The oxygen itself could be brought into the material by SiC dopant which can absorb oxygen and moisture during storage. There are two reasons why oxygen peak appears in the EDS spectrum within the grains. Firstly, it might be due to oxygen alloying in boron layers, which was reported to result in strong pinning in the MgB$_2$ thin films \cite{eom}. Secondly, the oxygen peak may be attributable to the presence of MgO nano-size inclusions formed within the grains.

\subsection{Effect of SiC-doping on $T_c$} \label{vol}

\begin{figure}
%\vspace{-0.5cm}
\includegraphics[scale=0.35]{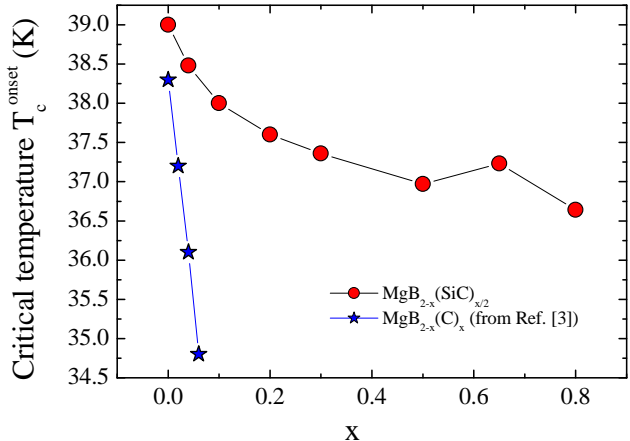}
\caption{\label{tcx} The critical temperature onset $T_c^{\rm onset}$ versus the doping level for all the measured samples. For comparison, $T_c$ for the single element doping taken from Ref.~\onlinecite{takenobu} is also shown.}
\end{figure}

Figure~\ref{tcx} shows the onset of critical temperature $T_c^{\rm onset}$ as a function of the doping level, as obtained by zero-field cooled magnetization measurements at $H = 25$~Oe. The onset was defined at $\chi_{dc} = - 5 \times 10^{-5}$~Emu/cm$^3$. For undoped sample, $T_c^{\rm onset} \simeq 38.98$~K. For the doped samples, $T_c^{\rm onset}$ decreases almost monotonically with increasing doping level. Surprisingly, the degradation of $T_c^{\rm onset}$ slows down, and likely reaches a quasi-plateau for $x \ge 0.5$, whereas for samples having $x \ge 1$ the bulk superconductivity vanishes, at least in the measured temperature range of $5~{\rm K} \le T \le 45$~K. The total $T_c^{\rm onset}$ drop is only 2.4~K for the superconducting samples with the SiC doping level up to $x = 0.8$ (nominal composition). In contrast, $T_c$ was depressed by almost 7~K in the case of pure C doping in MgB$_{2-x}$C$_x$ with $x = 0.2$ \cite{takenobu}; $T_c$ was reduced by about 0.5~K in the case of Si substitution in MgB$_{2-x}$Si$_x$ with $x = 0.05$ \cite{cimberle}. The latter result suggested that silicon substitution has much less dramatic effect on depression of $T_c$ compared to carbon. There is no data available for a higher Si doping level.

These results also support the idea of the counter-balance effect caused by C and Si co-substitution, producing the higher $T_c$ tolerance to MgB$_2$ structural changes induced by SiC doping. It is evident that the co-doping counter-balanced the negative effect on $T_c$ of the single element doping.

\begin{figure}
%\vspace{-0.5cm}
\includegraphics[scale=0.35]{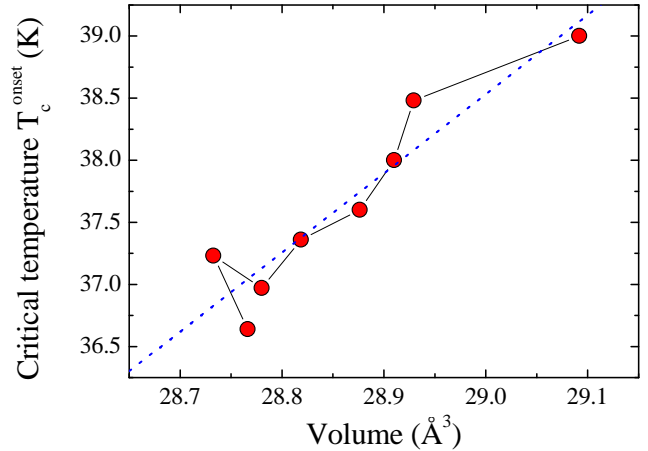}
\caption{\label{v}The crystal lattice unit cell volume change versus $T_c^{\rm onset}$ for the investigated doping range. The dotted line is a linear fit to the data.}
\end{figure}

As widely discussed in the literature, there are two factors: lattice structure change and electronic structure change that influence $T_c$ \cite{review}. From pressure experiments, it was found that $T_c$ decreases almost linearly with decreasing lattice volume (see the review paper \onlinecite{review} and references therein). As far as the lattice volume change is concerned, the C substitution for B would cause contraction of the crystal lattice and, hence, reduction of the unit cell volume; whereas the Si substitution for B would lead to expansion of the lattice and increase of the unit volume. Thus, it is plausible for MgB$_2$ to accommodate larger amounts of C and Si, than in the case of the single element doping. The larger level of substitution could bring forward an additional $T_c$ suppressing factor. Indeed, the reduction may occur not only due to the lattice contraction, but also due to alteration of the electron configuration with the doping. To assess each of the possible contributions, we have plotted $T_c^{\rm onset}$ versus crystal lattice unit cell volume as shown in Fig.~\ref{v}. The resulting dependence is rather linear. Similar linear dependences have been obtained for pressure experiments on pure MgB$_2$ material \cite{review}. Therefore, this similarity enables us to conclude that our co-substitution produces the same effect on $T_c$ as in the case of the pressure induced volume decrease, indicating that the volume (structural) factor plays the dominant role in affecting $T_c$ in Si and C co-doped MgB$_2$ system.
  
\subsection{Effect of SiC-doping on $J_c$ and pinning} \label{j}

\begin{figure}
%\vspace{-0.5cm}
\includegraphics[scale=0.33]{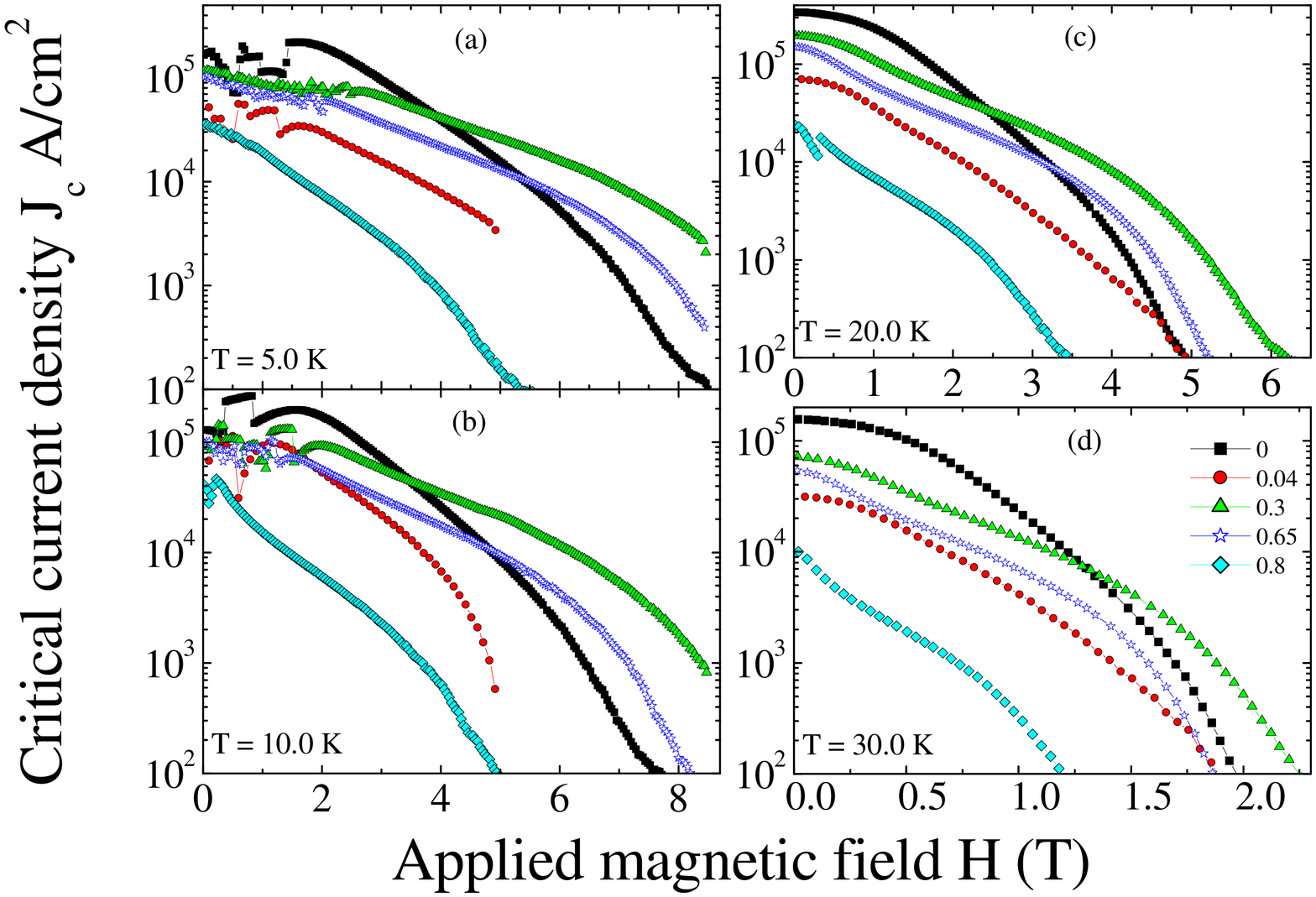}
\caption{\label{jc}Effect of SiC doping on $J_c(H)$ behavior of selected investigated samples compared to the pure MgB$_2$ sample at $T = 5$~K (a), 10~K (b), 20~K (c), and 30~K (d).}
\end{figure}

Fig.~\ref{jc} shows the $J_c(H)$ curves for selected samples at $T = 5$~K (a), 10~K (b), 20~K (c), and 30~K (d). The most striking feature shown in this figure is that $J_c(H)$-curves for samples with a wide range of doping cross the curve of the pure ($x = 0$) MgB$_2$ sample, exhibiting significantly higher critical current densities at higher applied fields. This is consistent with our first report on MgB$_2$(SiC)$_x$ samples \cite{sic}. The behavior is valid for the samples having the doping range of $0.1 \le x \le 0.65$, however $J_c(H)$-curves in Fig.~\ref{jc} are shown not for all measured samples for clarity. For example, samples in the doping range of $0.1 \le x \le 0.3$ have shown hardly distinguishable $J_c(H)$-behavior. It is important further to note at least two more interesting features revealed by our study. (i) The sample having $x = 0.65$, showed a relatively good $J_c(H)$ performance in high fields and $T < 30$~K (Fig.~\ref{jc}), but the striking is that it has as much as more than 50\% volume fraction of {\it non-superconducting phase}, according to the XRD results shown in Fig.~\ref{xrd}. (ii) At low doping level $x < 0.1$ the $J_c(H)$ behavior is surprisingly worse not only than that for the pure sample, but also for the higher doping level samples ($0.1 \le x < 0.8$). Below, we shall discuss possible reasons for the described $J_c(H)$ behavior shown in Fig.~\ref{jc}, and draw possible pinning mechanisms responsible for this behavior.

First, for our case we rule out the densification effect reported in the literature \cite{zhao,feng} for different doping materials, leading to a $J_c$ increase in rather low fields and temperatures. This would mean that, instead of introducing pinning into MgB$_2$, doping would help increase the density of the material, which would increase the effective cross-sectional area for supercurrent transport and reduce its percolative flow. In contrast to their works \cite{zhao,feng} but similarly to the work on Y$_2$O$_3$ nano-particle doping \cite{wang}, SiC doping showed no densification effect as evidenced by the fact that the density of our samples is independent of the doping level and equal to approximately 1.2~g/cm$^2$. This is less than 50\% of the theoretical density for this material, 2.63~g/cm$^2$. This can be understood because melting temperature of SiC is very high and SiC would not act as sintering aid in the temperature range of $800^{\circ}{\rm C}$ to 950$^{\circ}$C. Thus, the densification can not be claimed responsible for the improved $J_c(H)$ behavior at high fields in our doped samples.

In order to describe the origin of pinning enhancement which is responsible for the obtained $J_c(H)$ behavior, it is necessary to recognize the unique feature of SiC doping emphasized in the previous sections. It takes place in the form of the prevailing substitution and addition, while in works reported in the literature \cite{zhao,feng,wang,mickelson,takenobu,cimberle} element doping was in the form of additives, either not incorporated into the crystal lattice or incorporated to a lesser extent than in our work. This indicates that $J_c$ is very sensitive to the way impurities are introduced. A question arises, what kind of implications does the {\it substitution} have on pinning properties of the material?

\begin{figure}
%\vspace{-0.5cm}
\includegraphics[scale=0.37]{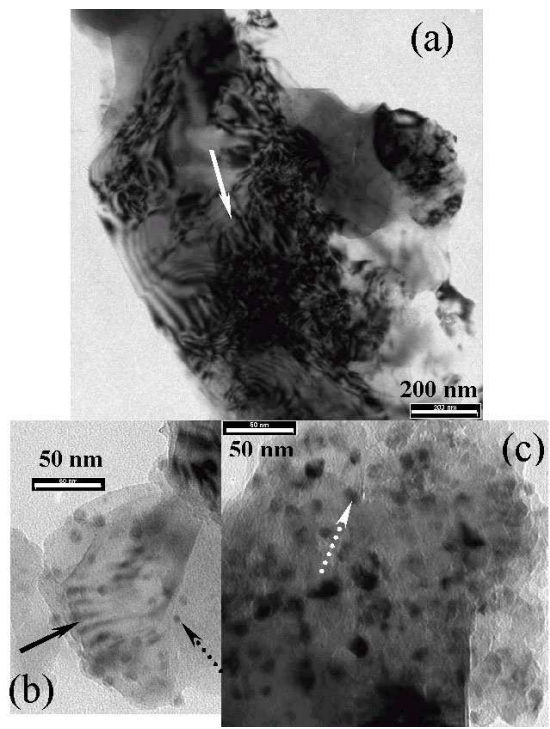}
%\vspace{0.5cm}
\caption{\label{tem}TEM images of the $x = 0.2$ and $x = 0.5$ samples. A high dislocation density within grains is shown in (a). Dislocations and round shape inclusions are present within each grain (b). The number of inclusions increases with the doping level (c). The solid and dotted arrows mark dislocations and inclusions, respectively.}
\end{figure}

As already shown in Fig.~\ref{cl}, the substitution causes the crystal lattice parameters to contract inducing local lattice strains in MgB$_2$-framework. This would create not only local fluctuation of the superconducting order parameter, but also appearance of a number of crystal defects. Indeed, our transmission electron microscopy (TEM) investigations clearly show a large number of dislocations within each MgB$_2$ grain (Fig.~\ref{tem}(a)), which are similar to those shown in our previous report \cite{sic}. Dislocations are known to serve as strong pinning centers \cite{pan}. Very limited structural study of the MgB$_2$ superconductor available in the literature indicated that pure MgB$_2$ has only very few atomic-scale dislocations \cite{hrem}, which can hardly provide strong pinning in high fields. Therefore, the high density of the dislocations induced by the substitution and revealed by TEM is a strong intra-granular source of pinning. The dislocation size is estimated to vary from 5~nm to 30~nm which, in average, is larger only by a factor of about 2 than the coherence length $\xi$ in MgB$_2$ superconductor \cite{review,finn}. This makes them perfect pinning sites, because each vortex line has the diameter of the normal core of $2\xi(T)$.

Another contributing source of pinning, as in the case of Y$_2$O$_3$ doping \cite{wang}, is non-superconducting nano-size precipitates, the amount of which grows with increasing doping level, as evident by the XRD patterns (Fig.~\ref{xrd}). Indeed, the TEM study shows 2~nm to 10~nm large round shape particles, presumably Mg$_2$Si, existing within each MgB$_2$ grain (Fig.~\ref{tem}(b,c)). Note, the inclusions are in average one order of magnitude smaller than initially added SiC-particles, indicating that the size of the initial dopant powder is not crucial for the final effect. Moreover, the amount of these inclusions increases with increasing doping level, in particular starting from $x \simeq 0.2$. At lower doping, fewer inclusion particles are observed. The optimal $J_c(H)$ dependence is reached for the sample with $x = 0.3$ (Fig.~\ref{jc}) when the volume fraction of the non-superconducting impurity Mg$_2$Si content was about 30\% (Fig.~\ref{xrd}). A further $x$-increase starts to worsen the $J_c(H)$ behavior, and, at the same time, the $a$-axis of the crystal lattice parameter exhibits a quasi-plateau within the experimental error (Fig.~\ref{cl}). This correlation points out that the optimal $J_c(H)$ behavior is reached when the substitution effect has reached its saturation. Presumably, because at this stage maximum amount of dislocations and rather high level of inclusions, balanced with a reasonable amount of superconducting phase co-exist. Further B replacement leads to the gradual elimination of the superconducting phase and, hence, worsening of $J_c(H)$ performance. The above arguments, together with the fact that inclusions alone \cite{wang} had a less pronounced effect than in our case, lead to a conclusion that at high fields the substitution, inducing a strongly developed net of dislocations, is the dominating source of the pinning in the material, which is possibly aided by the localized superconducting order parameter fluctuations. The secondary, however rather significant role, is held by the nano-size impurities confined within the grains. However, the point-like round shape inclusions seen in Fig.~\ref{tem}(b,c) can be effective only in the case of their extremely high density, since their pinning ability is of a ``collective" nature due to their point-like action. This influence would be less pronounced in high fields when the vortex-vortex interaction is very strong, which makes the vortex lattice more rigid, so that the point-like defects become less important. Since the thermally activated processes seem to be negligible in MgB$_2$ superconductor, it is likely that the point-like pinning would be temperature independent. This is in contrast to dislocations (extended defects) which pinning efficiency would be influenced by the temperature dependence of the coherence length. The higher the temperature, the shorter the coherence length, the weaker the pinning since the dislocation size remains unchanged. This behavior is seen in the experiment: at $T \le 20$~K the high-field part of $J_c(H)$-curves is much stronger for the best doping performance ($0.1 \le x \le 0.3$). This is an additional piece of the evidence to favor our scenario.

\begin{figure}
%\vspace{-0.5cm}
\includegraphics[scale=0.35]{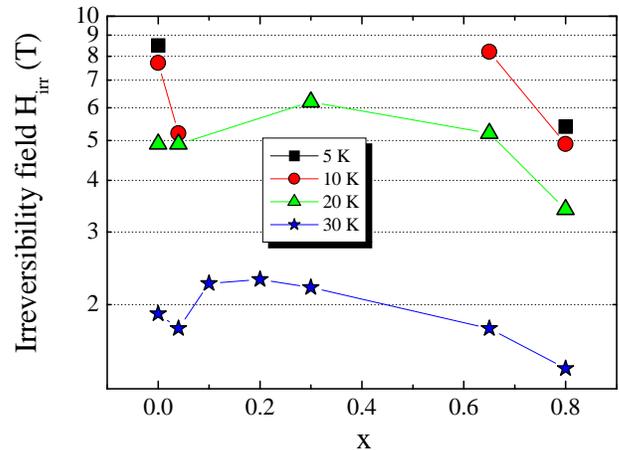}
\caption{\label{irr}The irreversibility field as a function of $x$ for different temperatures. At lower temperatures $T \le 10$~K, $H_{\rm irr}$ for the samples within $0.1 \le x \le 0.3$ doping range is increased beyond the field range accessible by our measuring equipment.}
\end{figure}

In figure~\ref{irr} the available irreversibility fields as a function of doping level are plotted for all the samples. Unfortunately, the upper field limitation in our experimental techniques did not allow us to measure $H_{\rm irr}$ for some samples at $T \le 10$~K, since the $H_{\rm irr}$ increase for these samples having doping level within $0.1 \le x \le 0.3$ was too large. However, we would like to note that the sample with $x = 0.65$ has $H_{\rm irr}$ larger than that of the pure sample at $T \le 20$~K; whereas at $T = 30$~K it becomes smaller than the corresponding value of the pure sample. This is in support of our temperature dependence of the pinning mechanism outlined above.

A puzzling situation is imposed by the lowest doping level sample having $x = 0.04$. It shows the $J_c(H)$ behavior which is worse than most of the other samples except the one with highest ``superconducting" doping level $x = 0.8$. Similar behavior was observed in our preceding work \cite{sic}, but it remained unexplained. This behavior is also well seen in Fig.~\ref{irr}. There is a minimum at $x = 0.04$, which is the easiest to notice for $T = 30$~K. We speculate that because of too-small level of substitution, neither sufficiently large density of dislocations nor inclusions could be formed. On the other hand, the doping did cause a significant crystal lattice contraction, which might have lead to strongly distorted super-current paths (for example due to a higher level of porosity) leading to the poor $J_c(H)$ behavior observed.

\section{Conclusion}

The main result of this work is the introduction of atomic substitutions of both Si and C in the crystal lattice of the MgB$_2$ superconductor. This was possible due to the counter-balanced atomic size of Si and C, in average nearly coinciding with the size of B. The substitution was shown to prevail up to rather high doping level $x \simeq 0.3$ in the MgB$_{2-x}$(SiC)$_{x/2}$ composition. At higher doping level the doping results in impurities, which eventually suppress the bulk superconductivity. The substitution was shown to very strongly improve $J_c(H)$ dependence at high applied fields. We suggested a pinning mechanism explaining this improvement, in which the main intra-granular ingredients are the dominating contribution of a large number of dislocations induced by substitution effect and the secondary contribution of the nano-inclusions.
 
\begin{acknowledgments}
We thank E. W. Collings, R. Neal, T. Silver, M.J. Qin, M. Sumption, and M. Tomsic for their helpful discussions. This work was supported by the Australian Research Council, Hyper Tech Research Inc OH USA, Alphatech International Ltd, NZ, and the University of Wollongong.
\end{acknowledgments}

%\bibliography{prlmgb2}

\begin{thebibliography}{99}
\bibitem{nature1} J. Nagamatsu, N. Nakagawa, T. Muranaka, Y. Zenitani, and J. Akimitsu, Nature {\bf 410}, 63 (2001).
\bibitem{review}C. Buzea and T. Yamashita, Supercond. Sci. Technol., {\bf 14}, R115 (2001). 
\bibitem{takenobu}T. Takenobu, T. Ito, D.H. Chi, K. Prassides, and Y. Iwasa, Phys. Rev B {\bf 64}, 134513 (2001).
\bibitem{mickelson}W. Mickelson, J. Cumings, W.Q. Han, and A. Zettl, Phys. Rev. B {\bf 65}, 052505 (2002).
\bibitem{cimberle}M.R. Cimberle, M. Novak, P. Manfrinetti, and A. Palenzona, Supercond. Sci. Tech. {\bf 15}, 43 (2002).
\bibitem{goldacher}W. Goldacher, S. I. Schlachter, S. Zimmer, and H. Reiner, Supercond. Sci. Technol. {\bf 14}, 787 (2001).
\bibitem{suo} H. L. Suo, C. Beneduce, M. Dhalle, N. Musolino, J. Y. Genoud, and R. Flukiger, Appl. Phys. Lett. {\bf 79}, 3116 (2001).
\bibitem{grasso}G. Grasso, A. Malagoli, C. Ferdeghini, S. Roncallo, V. Braccini, M. R. Cimberle, and A.S. Siri, Appl. Phys. Lett. {\bf 79}, 230 (2001).
\bibitem{glowacki}B. A. Glowacki, M. Majoros, M. Vickers, J. E. Evetts, Y. Shi, and I. McDougall, Supercond. Sci. Technol. {\bf 14}, 193 (2001)
\bibitem{jin}S. Jin, H. Mavoori, and R. B. van Dover, Nature {\bf 411}, 563 (2001).
\bibitem{soltan}S. Soltanian, X. L. Wang, I. Kusevic, E. Babic, A. H. Li, M. J. Qin, J. Horvat, H. K. Liu, E. W. Collngs, E. Lee, M. D. Sumption, and S. X. Dou, Physica C {\bf 361}, 84 (2001).
\bibitem{takano}Y. Takano, H. Takeya, H. Fujii, H. Kumakura, T. Hatano, K. Togano, H. Kito, and H. Ihara, Appl. Phys. Lett. {\bf 78}, 2914 (2001).
\bibitem{eom}C. B. Eom, M. K. Lee, J. H. Choi, L. Belenky, X. Song, L. D. Cooley, M. T. Naus, S. Patnaik, J. Jiang, M. Rikel, A. Polyanskii, A. Gurevich, X. Y. Cai, S. D. Bu, S. E. Babcock, E. E. Hellstrom, D. C. Larbalestier, N. Rogado, K. A. Regan, M. A. Hayward, T. He, J. S. Slusky, K. Inumaru, M. K. Haas, R. J. Cava, Nature {\bf 411}, 558 (2001).
\bibitem{bogu}Y. Bugoslavsky, L. F. Cohen, G. K. Perkins, M. Polichetti, T. J. Tate, R. G. William, and A. D. Caplin, Nature {\bf 411}, 561 (2001).
\bibitem{zhao}Y. Zhao, Y. Feng, C. H. Cheng, L. Zhou, Y. Wu, T. Machi, Y. Fudamoto, N. Koshizuka, and M. Murakami, Appl. Phys. Lett. {\bf 79}, 1154 (2001).
\bibitem{feng}Y. Feng, Y. Zhao, Y. P. Sun, F. C. Liu, B. Q. Fu, L. Zhou, C. H. Cheng, N. Koshizuka, and M. Murakami, Appl. Phys. Lett. {\bf 79}, 3983 (2001).
\bibitem{wang}J. Wang, Y. Bugoslavsky, A. Berenov, L. Cowey, A. D. Caplin, L. F. Cohen, J. L. M. Driscoll, Cond-mat/0204472.
\bibitem{nature2} D. C. Larbalestier, M. O. Rikel, L. D. Cooley, A. A. Polynaskil, J. Y. Jiang, S. Patniak, X. Y. Cai, D. M. Feldman, A. Gurevich, A. A. Squitieri, M. T. Naus, C. B. Eom, E. E. Hellstrom, R. J. Cava, K. A. Regan, N. Rogado, M. A. Hayward, T. He, J. S. Slisky, P. Khalifah, K. Inumara, and M. Haas, Nature {\bf 410}, 186(2001).
\bibitem{dou}S. X. Dou, X. L. Wang, J. Horvat, D. Milliken, A. H. Li, K. Konstantinov, E. W. Collings, M. D. Sumption, and H. K. Liu, Physica C {\bf 361}, 79 (2001).
\bibitem{sic}S. X. Dou, A. V. Pan, S. Zhou, M. Ionescu, H. K. Liu, and P. R. Munroe, Cond-mat/0206444.
\bibitem{wang2} X .L. Wang, S. Soltanian, J. Horvat, M. J. Qin, H. K. Liu, and S. X. Dou, Physica C {\bf 361}, 149 (2001).
\bibitem{pan}See for example V. M. Pan and A. V. Pan, Fiz. Nizk. Temp. {\bf 27}, 991 (2001) [(Low Temp. Phys. {\bf 27}, 732 (2001)] and references therein. \bibitem{hrem}Y. X. Chen, D. X. Li, and G. D. Zhang, Mater. Sci. Eng. A (2002), in press.
\bibitem{finn}D. K. Finnemore, J. E. Ostenson, S. L. Bud'ko, G. Lapertot, and P. C. Canfield, Phys. Rev. Lett. {\bf 86}, 2420 (2001).

\end{thebibliography}

\end{document}